\documentclass[letterpaper, 10 pt, conference]{ieeeconf} 
\usepackage{graphicx}                                                      
\usepackage{amsmath} 
\usepackage{amssymb}
\usepackage{pstricks}
\usepackage{pst-all}
\usepackage{pst-plot}
\usepackage{pstricks-add}
\usepackage{pst-math}
\usepackage{pst-func}
\usepackage{url}
\usepackage{amsmath}
\IEEEoverridecommandlockouts                              
\usepackage[utf8]{inputenc} 
\usepackage{amsmath,amssymb,epsfig}

\usepackage{graphicx}
\usepackage{url}

\title{Verifying Response Times in Networked Automation Systems Using Jitter Bounds}

\author{Seshadhri Srinivasan$^{1}$, Furio Bounapane $^{2}$, J\"{u}ri Vain$^{2}$,  and Srini Ramaswamy$^{3}$
\thanks{This work was supported by Group of Researchers in Automatic Control Engineering (GRACE)}
\thanks{$^{1}$ S. Seshadhri is with the  Dept. of Engineering, University of Sannio, Benevento Italy}
\thanks{$^{4}$ Furio Buonopane is with Engineering Department (DICEA), University of Naples Federico II, Naples, Italy, 80125
e-mail: furio.buonopane@gmail.com}
\thanks{$^{3}$Srini Ramaswamy is with India Corporate Research Center, ABB Global Industries and Services Ltd., Bangalore, India 
srini.ramaswamy@in.abb.com}
\thanks{$^{2}$ J\"{u}ri Vain is with the Institute of Cybernetics at Tallinn University of Technology, Estonia}
}

\begin{document}
\maketitle
\thispagestyle{empty}
\pagestyle{empty}

\begin{abstract}
Networked Automation Systems (NAS) have to meet stringent response time during operation. Verifying response time of automation is an important step during design phase before deployment. Timing discrepancies due to hardware, software and communication components of NAS affect the response time. This investigation uses model templates for verifying the response time in NAS. First, jitter bounds model the timing fluctuations of NAS components. These jitter bounds are the inputs to model templates that are formal models of timing fluctuations. The model templates are atomic action patterns composed of three composition operators- sequential, alternative, and parallel and embedded in time wrapper that specifies clock driven activation conditions. Model templates in conjunction with formal model of technical process offer an easier way to verify the response time. The investigation demonstrates the proposed verification method using an industrial steam boiler with typical NAS components in plant floor.

\end{abstract}

\IEEEpeerreviewmaketitle
\section{Introduction}
Networked automation systems (NAS) in industrial automation refer to systems with networked sensors, actuators and controllers \cite{Vogel2012}. Response time is defined as the difference between the cause of an event (a new sensor measurement) to the effect on the technical process. Industrial automation systems are real-time systems requiring fast response times (typically in milli-seconds). Response time in NAS need to be verified during design phase to avoid re-design after deployment. The importance of response time in NAS is demonstrated from the numerous approaches proposed in literature (see, \cite{Pereira2004}-\cite{Addad2010} and references therein). Computing the bound of timing fluctuations remains the focus of these approaches. Numerous applications of NAS has been reported in literature (see,\cite{SeshCC}-\cite{Ganesh2014} and references therein).

Timed model-checking is a promising technique to analyse critical systems, because it performs exhaustive checking using formal models. Recent research uses tools from model-checking to verify the response time in NAS. To our best knowledge, Frey et al. \cite{Frey2005} were the first to use model-checking tools in NAS to study component failures using probabilistic model checking (PMC) without proposing a formal model. Later, the authors studied the simulation of response time using Dymola/Modelica NAS component models in \cite{Frey2007}. This method suffers from the limitations of simulation i.e. to test specific scenarios against exhaustive verification offered by model checking. The model-checking methods proposed in \cite{Mazz2010} and \cite{Ruel2009} lack the support of modelling framework and therefore, are restricted to specific architecture or scenarios. Vogel-Heuser et al. \cite{Vogel2011} presented a component oriented modelling approach that captures the timing requirements and specifications to verify response time in NAS.

A reading of the literature reveals that modelling the timing fluctuations due to communication, physical and software components of NAS and their time-variations offer stiff challenge to verify the timing performance using timed-model checking. To overcome these challenges, this investigation uses NAS component models capturing the timing fluctuations as jitter. Composition of these components using the time-chain model in \cite{Vogel2011} with additional specification on jitter bounds, and nature of their variation (termed as behaviour) leads to the jitter time-chain. The jitter time-chain is used to create model templates of NAS components for verifying response times. The model templates are atomic action patterns with three composition operators to model the jitter. The model templates of jitter in conjunction with the formal model of the process defines the formal model required for verifying the response time. The use of model template simplifies the procedure for generating formal model useful for verifying response time of NAS. 

The main contributions of this investigation are- jitter based model for verifying the timing performance of NAS, model patterns that use jitter bounds to model the timing imperfections, and illustration of the verification procedure using steam boiler in industrial plant-floor. The paper has five sections including the introduction. Section II, presents the jitter based timing model of NAS and the discussion on model patterns is in section III. Example in section IV illustrates the use of model templates, and section V presents the conclusion of the investigation.

\section{Modelling Timing Imperfections in NAS}
The timing imperfections in NAS are due to hardware, software, and communication components. Hardware timing imperfections are due to sensors, actuators, signal processing, and controller hardware. Fig. 1 shows the sources of hardware jitter in NAS. The hardware jitter is modelled to be constant as the variation happens over long-time frames. Software jitter are mainly due to scheduling, cache memory, pre-emption, interrupts, context switching, dynamic control algorithms, multiple loops and asynchronous communication between tasks. As software timing imperfections are usually measured using execution times (such as best-case execution time, worst-case execution time, average execution time etc.) they naturally suggest the use of deterministic model.

The timing imperfections in NAS can be broadly classified into three broad categories, they are: {\em{(i)}} hardware, {\em{(ii)}} software, and {\em{(iii)}} network- induced \cite{SeshThe,SeshCC}. Model of the timing discrepancies is required to verify the response time using formal models. Based on the timing imperfections the delay time-chain can be drawn as shown in Fig. \ref{fig:timechain}.

\begin{figure}[h]
\centering
\includegraphics[width=8cm,height=8cm]{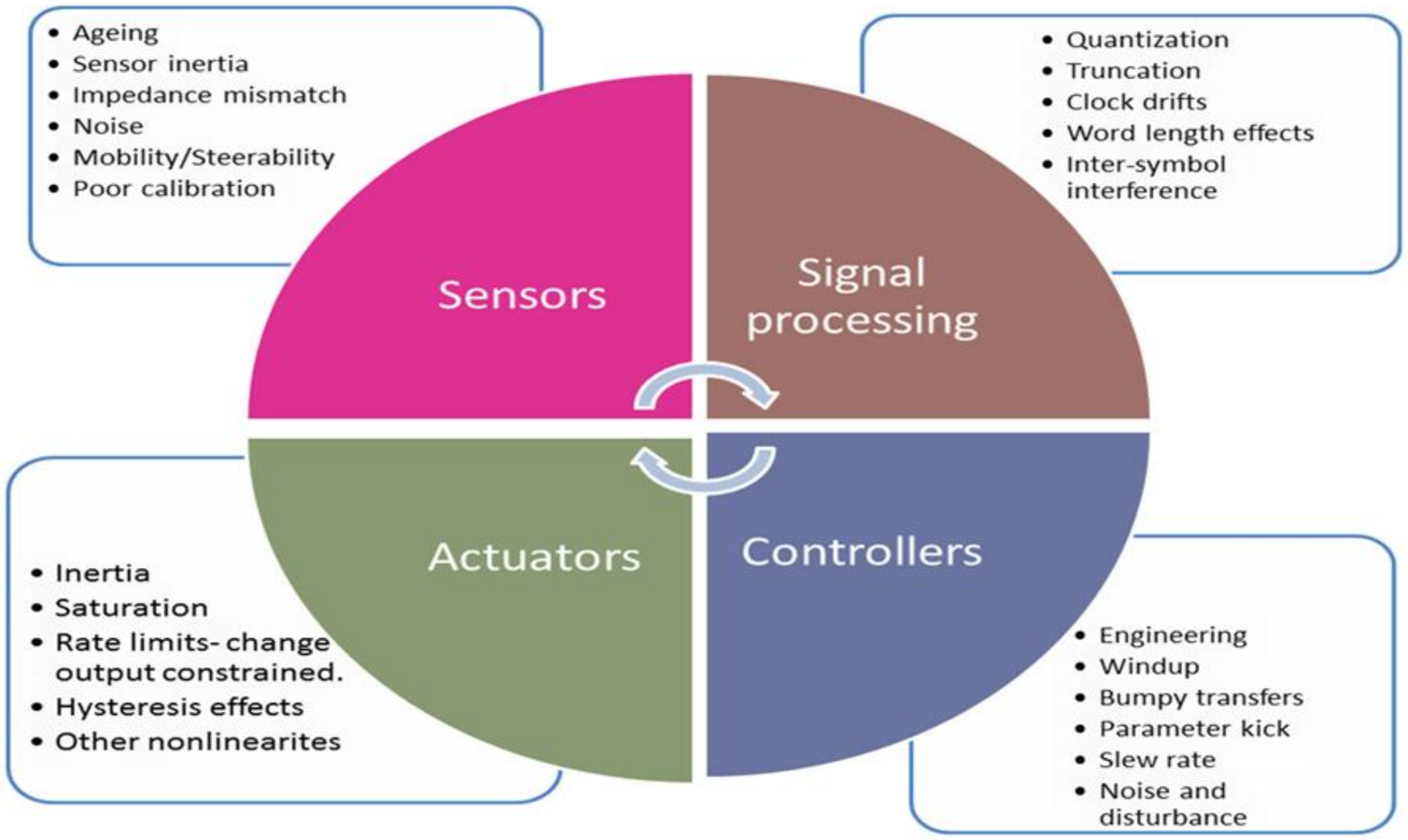}
\caption{Hardware timing jitter}
\label{fig:hardjitter}
\end{figure}

\begin{figure}[h]
\centering
\includegraphics[scale=0.3]{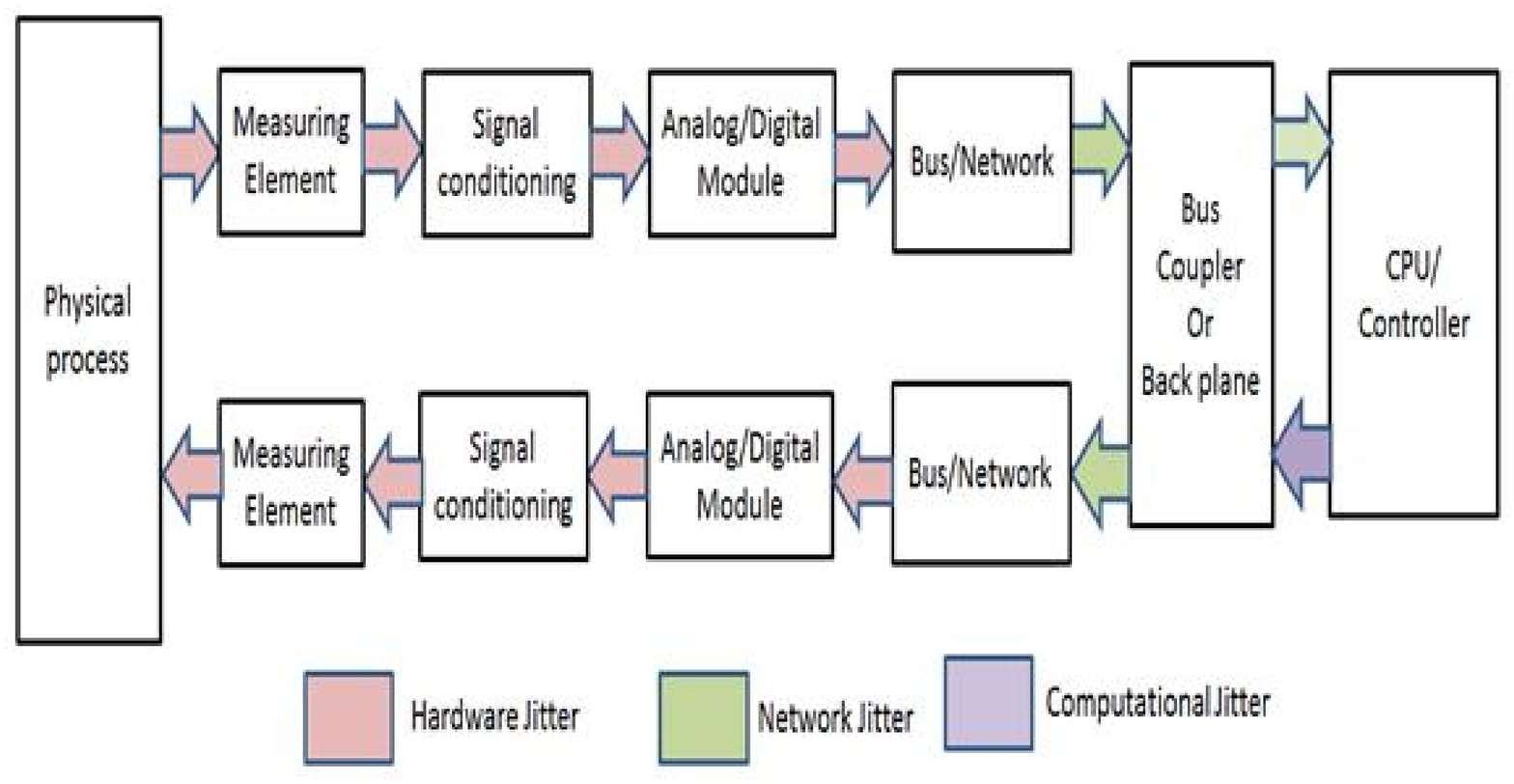}
\caption{Time-Chain for timing performance specification}
\label{fig:timechain}
\end{figure}

Hardware timing imperfections are due to sensors, actuators, signal processing, and controller hardware. The sources of hardware jitter in NAS due to NAS components is shown in Fig. \ref{fig:hardjitter}. This investigation proposes to model the hardware jitter as constant, as usually the timing imperfections are found to vary over long time-frames. Software jitter are mainly due to scheduling, cache memory, pre-emption, interrupts, context switching, dynamic control algorithms, multiple loops and asynchronous communication between tasks. As software timing imperfections are usually measured using execution times (such as best-case execution time, worst-case execution time, average execution time etc.) they naturally suggest the use of deterministic model.

Communication related timing imperfections include latencies, and data-loss. Latencies in communication channels depend on many parameter such as length of communication channel, channel load, protocol employed, network interface card employed in automation, and contention ratio. As these parameters are inherently random, they make communication latencies time-varying and many models have been proposed for modeling time-varying delays (see, \cite{SeshThe}-\cite{Nilson}). The timing imperfections are modeled using jitter bounds on individual components. Communication jitter is modeled using to be time-varying but bounded as
\begin{equation}
J_C \in [J_C^{min}, J_C^{max}]
\label{eq:jitterCom}
\end{equation}
where $J_C$ is the total communication jitter, $J_C^{min}$ and $J_C^{max}$ are the minimum and maximum communication jitter, respectively. Therefore, the total jitter in NAS is
\begin{equation}
J_T =J_H+J_S+J_C
\label{eq:jitterCom}
\end{equation}
where $J_T$ is the total jitter in the NAS, $J_H$, and $J_S$ are the hardware, software, and communication jitter, respectively. 




\section{Model Patterns for Response Time Verification}
\begin{figure}[h]
\centering
\includegraphics[scale=0.32]{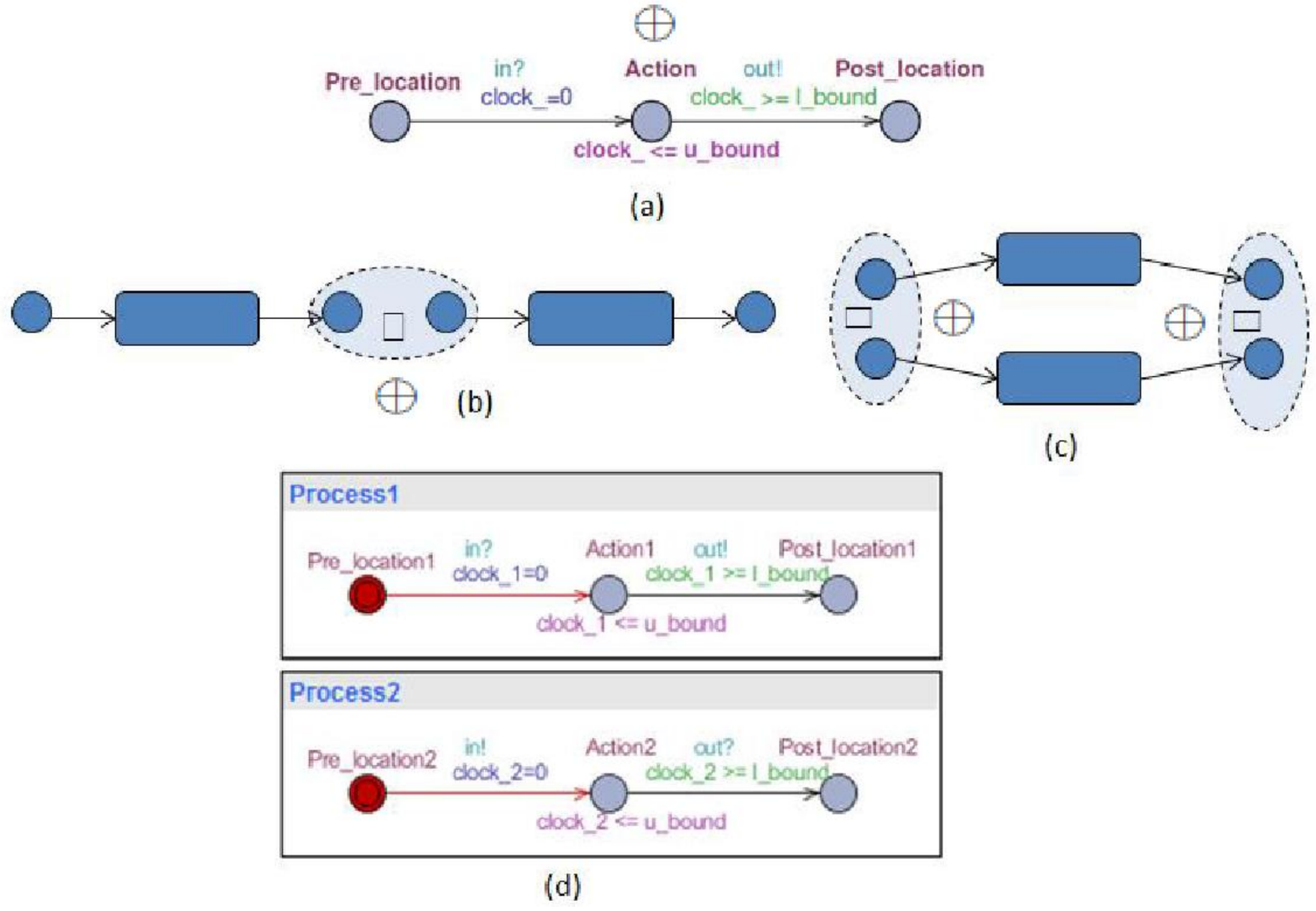}
\caption{Delay model patterns}
\label{fig:ModelPattern}
\end{figure}

Having obtained the model of NAS by composing components, the time chain can be generated as a formal model using model patterns that take the jitter bounds as inputs. To construct the time-chain, this investigation assumes two types of timing performance components:
\begin{itemize}
	\item components that are activated by some external event using model patterns (see, model patterns in Fig. \ref{fig:ModelPattern})
	\item components are activated periodically each with possibly different period and jitter
\end{itemize}
To model timing fluctuations this investigation proposes structural modeling approach that considers the models that are constructed from an atomic action pattern (Figure 4(a)) by means of three composition operators: sequential (Fig. \ref{fig:ModelPattern}(b)), alternative  (Fig. \ref{fig:ModelPattern} (c)), and parallel composition ((Fig. \ref{fig:ModelPattern} (d)). For parallel composition an additional channel matching constraint is required: when ever there is a synchronization condition in one of the parallel components then there must be also matching synchronization condition in the other parallel component. We call the models constructed that way well-formed models. Atomic action model pattern captures the lower and upper bound as $[l_{bound}, u_{bound}]$ as shown in Fig. \ref{fig:ModelPattern} (a). This model pattern for delay is particularly useful in scenarios requiring action triggered by an external event (induced by another component). The communication jitter on a component can be modeled using interval characterization of the jitter. 

Sequential and alternative compositions are defined as applications of location merging operator $\oplus$ on two well-formed timed automata.
\begin{equation}
Post_1 \oplus Pre_2
\label{eq:SeqComp}
\end{equation}
where $Post_1$ and $Pre_2$ indicate respectively the Post- and Pre-locations of the first and second component automata post-conditions of the automata (see, Fig. \ref{fig:ModelPattern} (b). This can be used in scenarios wherein one timing imperfection due to one component results in timing imperfection in other component. 

The other composition model pattern is the alternative composition shown in Fig. \ref{fig:ModelPattern} (c) and the result of the model pattern is given by
\begin{equation}
Pre_1+Pre_2  \vee Post_1+Post_2  \vee (Pre_1+Pre_2,Post_1,Post_2)
\label{eq:AltComp}
\end{equation}

In parallel composition ($\parallel$) shown in Fig. \ref{fig:ModelPattern} (d) the $in$ and $out$ indicate the channel name suffixes $\!$, and $?$ indicate the  synchronization direction.  The model patterns of Fig. \ref{fig:ModelPattern} give the formal models for capturing jitter in NAS along with the physical components that are triggered by an external event (synchronization condition "in?" in Fig.  \ref{fig:ModelPattern}(a)).

\subsection{Timing Wrapper}
\begin{figure}[h]
\centering
\includegraphics[scale=0.2]{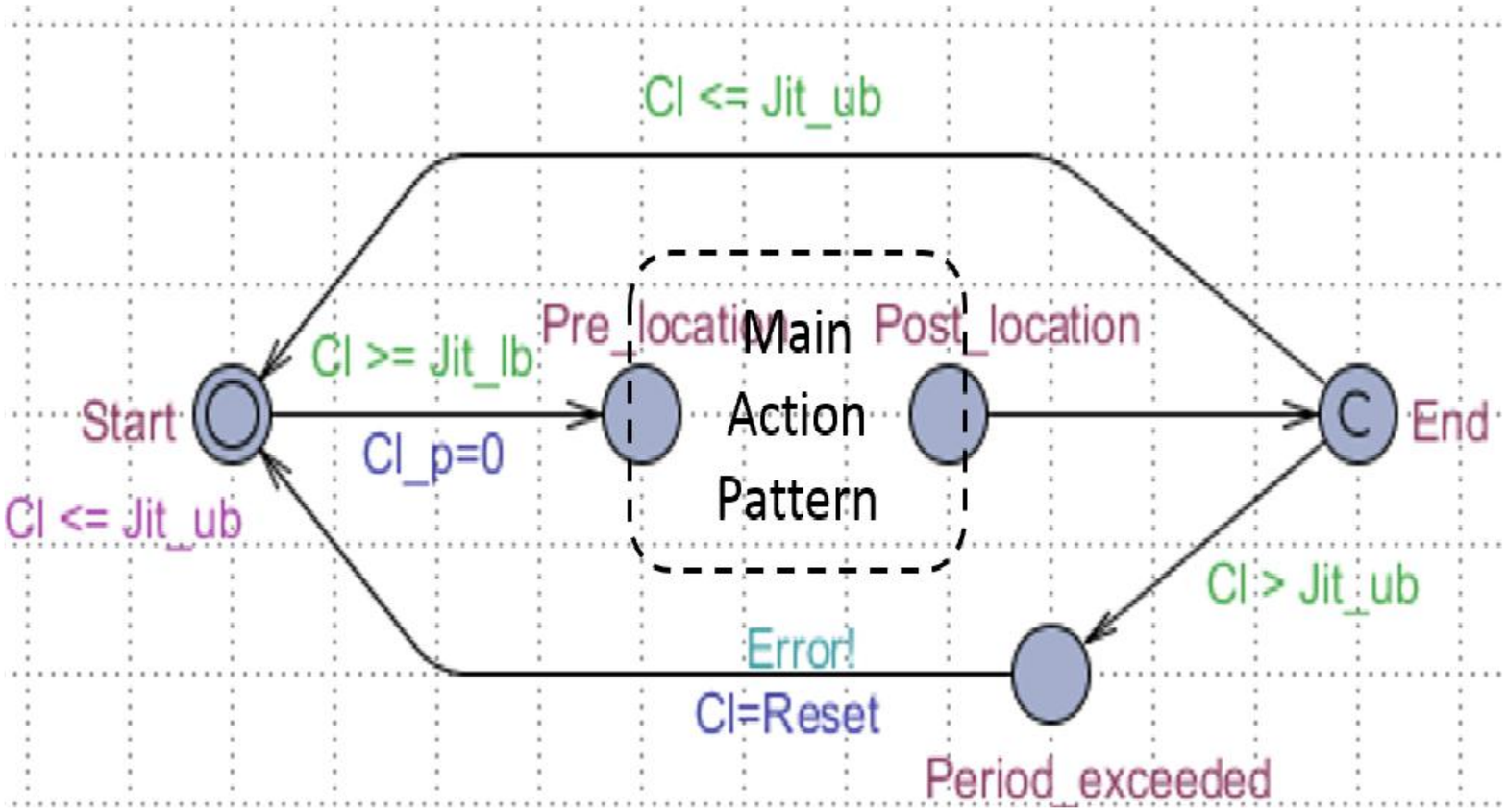}
\caption{Timing Wrapper}
\label{fig:TimeWrap}
\end{figure}

For modeling the periodically triggered (by clock) actions the construct timing wrapper is introduced in addition to main model patterns described in Subsection A.  (Due to the  limited space other clock triggered activation patters  implementable in Timing Wrapper are not considered in this paper. ) timing wrapper introduces an auxillary clock $Cl$ that is needed for modelling the activation period with jitter within the interval $[Jit_{lb},Jit_{ub}]$ as shown in Fig. \ref{fig:TimeWrap}.

The model patterns described in this section along with the timing wrapper can be used to capture the timing imperfections as jitter in the formal model of NAS along with that of the technical process. Thus timing model of the time-chain (see, Fig. \ref{fig:timechain}) for timing performance specification can be constructed from patterns 1-4 depending on the way of activation of each component in the time chain. 

The next step of the work-flow consists of simulation of both the process and the timing components to identify critical operating points of the technical process and the test conditions for the NAS timing components. Simulation of technical process can be used to identify the critical points that needs to be tested. The following example illustrates the use of simulation to study the timing performance.

\section{Results}
This section presents an example of using the work-flow for timed-model checking. The example considered is a steam boiler.
\subsection{Description of the Technical Process}
\begin{figure}[h]
\centering
\includegraphics[scale=0.2]{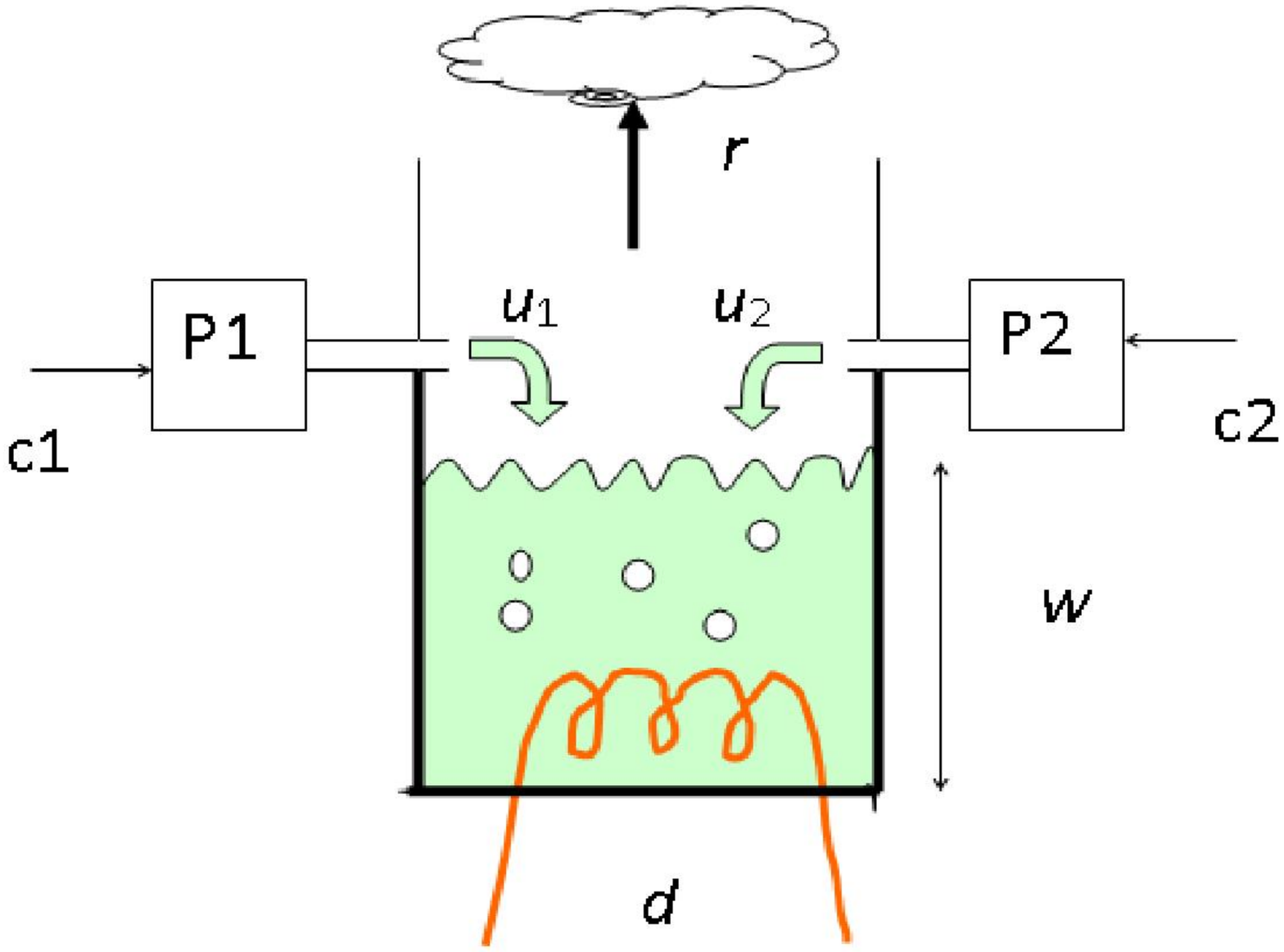}
\caption{Steam Boiler}
\label{Fig:SteamBoiler}
\end{figure}

The steam boiler consists of two pumps $P_1$ and $P_2$ and heater as shown in Fig. \ref{Fig:SteamBoiler}. Here, $w$, $u_1(t)$, $u_2(t)$, $d$ denote the water-level of the boiler ($w~>0$), inflow of pump~1 in $l/min$, inflow of pump~2 in $l/min$, and the power of the boiler. The vaporization ratio is denoted using $\dot{r}$. 

\noindent{\em{Assumptions:}}\\
At each point of time $t$ pump $P_i$ either is working ($u_i(t) = P_i$) or is stopped ($u_i(t) = 0$). There is delay $T_i$ between  $i$-th switching on and when the pump starts actually pumping. There is no such a delay when the pump is switched off.

\begin{figure}[h]
\centering
\includegraphics[scale=0.2]{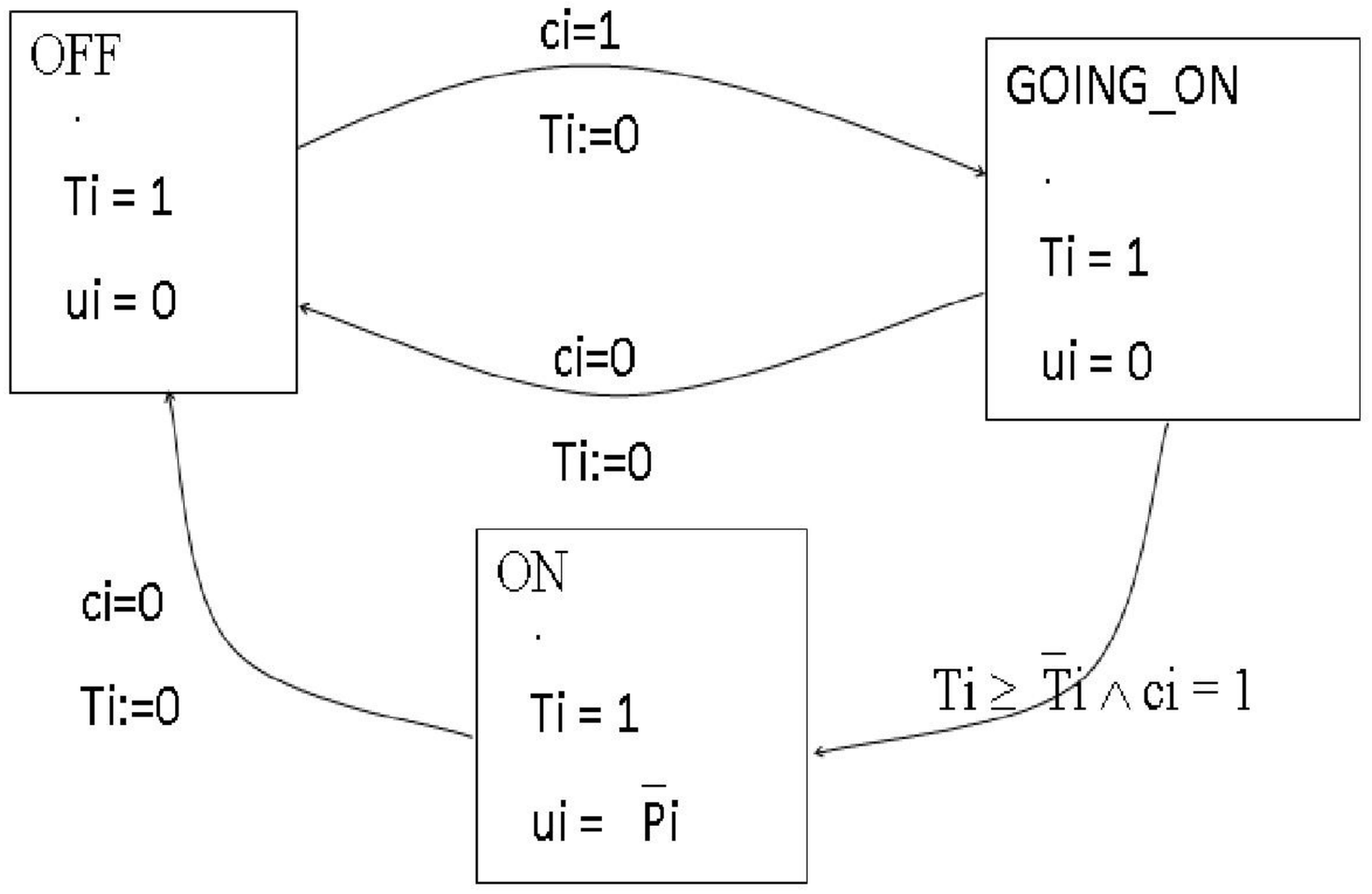}
\caption{Hybrid model of the boiler}
\label{Fig:HybBoiler}
\end{figure}

The working of steam boiler can be described using the hybrid automata in Fig. \ref{Fig:HybBoiler}

\begin{figure}[h]
\centering
\includegraphics[scale=0.32]{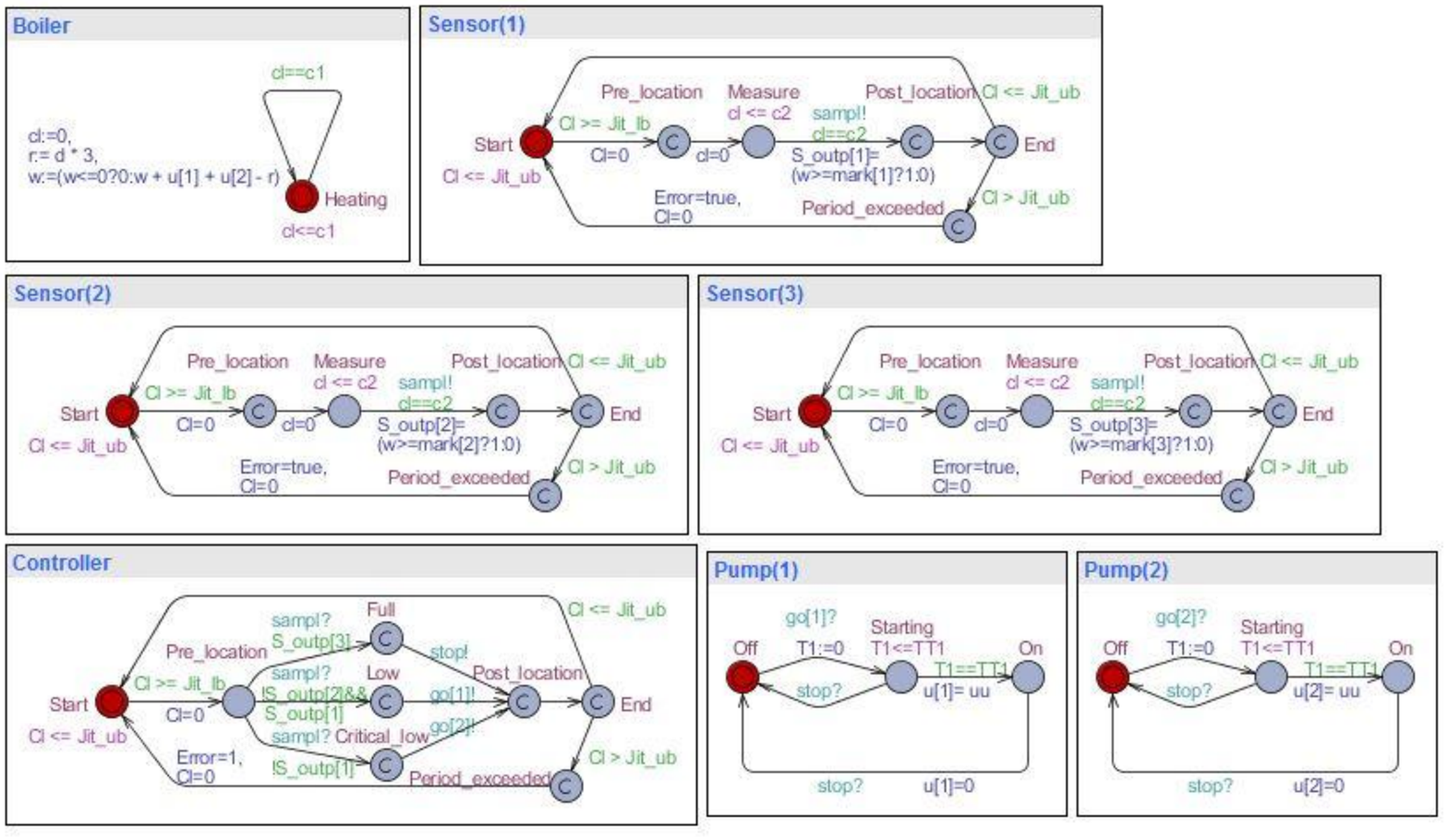}
\caption{UPPAAL formal model of the steam boiler with model templates}
\label{Fig:result1}
\end{figure}

The model-templates using action model patterns and composition operators can be used to construct the formal model of the timing fluctuations, and time-wrapper can be used in case of periodic operations. This formal model can be composed with the formal model of the components of the steam boiler that could be used for verifying the response times of NAS. These formal models are modeled in UPPAAL as timed-automata \cite{Beng2004} models  shown in Fig. \ref{Fig:result1}. The reaction time verification on given model is implemented by a model checking query that uses standard TCTL logic operator "time bounded leads to", i.e.,  Stimulus $\rightarrow_d$ Response, where Stimulus and Response are 1st order state formuli that specify the begin and end events of the reaction time bound $d$ to be verified.

\section{Conclusion}
This paper presented a simulation driven verification work-flow for verifying the response time in NAS. The approach modelled the timing discrepancies in the NAS components using jitter bounds based on their occurrence as constant, deterministic and time-varying. Obtained jitter bounds were used to generate model templates considering various scenarios that arise in NAS. Then simulation is done on the technical process along with knowledge of jitter to obtain results useful for model abstraction and verification. The inputs from the modelling, template generation, and simulation steps are used to verify the response time of the NAS. The work-flow was illustrated using a plant-floor example of steam boiler and pH neutralization process. Extending the work-flow to verify other timing properties and extending to verify multi-core automation systems are future course of this investigation.

\end{document}